\pdfoutput=1

\documentclass[pdflatex,sn-basic,iicol]{sn-jnl}

\jyear{2022}%

\usepackage{dblfloatfix} 
\usepackage{natbib}
\begin{document}

\title[" "]{Neural manifold analysis of brain circuit dynamics in health and disease}


\author*[1,2]{\fnm{Rufus} \sur{Mitchell-Heggs}}\email{rmitch11@ed.ac.uk}
\equalcont{These authors contributed equally to this work.}

\author[1,3]{\fnm{Seigfred} \sur{Prado}}
\equalcont{These authors contributed equally to this work.}

\author[1]{\fnm{Giuseppe P.} \sur{Gava}}
\equalcont{These authors contributed equally to this work.}

\author[1]{\fnm{Mary Ann} \sur{Go}}

\author*[1]{\fnm{Simon R.} \sur{Schultz}}\email{s.schultz@imperial.ac.uk}

\affil*[1]{\orgdiv{Department of Bioengineering and Centre for Neurotechnology}, \orgname{Imperial College London}, \orgaddress{\city{London}, \postcode{SW7 2AZ}, \country{United Kingdom}}}

\affil[2]{\orgdiv{Centre for Discovery Brain Sciences}, \orgname{The University of Edinburgh}, \orgaddress{\city{Edinburgh}, \postcode{EH8 9XD}, \country{United Kingdom}}}

\affil[3]{\orgdiv{Department of Electronics Engineering}, \orgname{University of Santo Tomas}, \orgaddress{\city{Manila}, \country{Philippines}}}

\abstract{
Recent developments in experimental neuroscience make it possible to simultaneously record the activity of thousands of neurons. However, the development of analysis approaches for such large-scale neural recordings have been slower than those applicable to single-cell experiments. One approach that has gained recent popularity is neural manifold learning. This approach takes advantage of the fact that often, even though neural datasets may be very high dimensional, the dynamics of neural activity tends to traverse a much lower-dimensional space. The topological structures formed by these low-dimensional neural subspaces are referred to as ``neural manifolds", and may potentially provide insight linking neural circuit dynamics with cognitive function and behavioral performance. In this paper we review a number of linear and non-linear approaches to neural manifold learning, including principal component analysis (PCA), multi-dimensional scaling (MDS), Isomap, locally linear embedding (LLE), Laplacian eigenmaps (LEM), t-SNE, and uniform manifold approximation and projection (UMAP). We outline these methods under a common mathematical nomenclature, and compare their advantages and disadvantages with respect to their use for neural data analysis. We apply them to a number of datasets from published literature, comparing the manifolds that result from their application to hippocampal place cells, motor cortical neurons during a reaching task, and prefrontal cortical neurons during a multi-behavior task. We find that in many circumstances linear algorithms produce similar results to non-linear methods, although in particular cases where the behavioral complexity is greater, non-linear methods tend to find lower-dimensional manifolds, at the possible expense of interpretability. We demonstrate that these methods are applicable to the study of neurological disorders through simulation of a mouse model of Alzheimer's Disease, and speculate that neural manifold analysis may help us to understand the circuit-level consequences of molecular and cellular neuropathology.
}




\maketitle

\section{Introduction}\label{sec1}

While the investigation of single neurons has undoubtedly told us much about brain function, it is uncertain whether individual neuron properties alone are sufficient for understanding the neurobiological basis of behavior \citep{pang2016dimensionality}. In some cases, trial-averaging of single-neuron responses may lead to confusing or misleading interpretation of true biological mechanisms \citep{sanger2014crouching, cunningham2014dimensionality}. Additionally, single-neuron activities studied in higher-level brain areas involved in cognitive tasks \citep{machens2010functional, laurent2002olfactory, churchland2010cortical} are highly heterogeneous both across neurons and across experimental conditions even for nominally identical trials. And finally, it may well be that task-relevant information is represented in patterns of activity across multiple neurons, above and beyond what is observable at the single neuron level. Unfortunately, characterizing such patterns, in the worst case, may require measurement of an exponential number of parameters (the ``curse of dimensionality").

However, it appears that in many circumstances, patterns of neural population activity may be described in terms of far fewer population-level features than either the number of possible patterns, or even the number of neurons observed \citep{Churchland2012NeuralReaching,mante2013context,Chaudhuri2019TheSleep,Gallego2020Long-termBehavior,nieh2021geometry,stringer2019high}. This underpins a paradigm shift in the studies of neural systems from single-neuron analysis to a more comprehensive analysis that integrates single-neuron models with neural population analysis. 

If, as appears to be the case, the spatiotemporal dynamics of brain activity is low dimensional, or at least much lower-dimensional than the pattern space, then it stands to reason that such activity can be characterized without falling afoul of the curse of dimensionality, and on reasonable experimental timescales. Indeed, in recent years, numerous techniques have been developed to do just that. For instance, classification algorithms have been applied  to neuronal ensembles to predict aspects of behavior \citep{rigotti2013importance,fusi2016neurons,rust201430,raposo2014category}. One problem with this is that the common practice is to identify “neuronal ensembles” by grouping together neurons with sufficiently highly correlated activity during the same behaviors or in response to the same stimuli. This ignores information that is transmitted collectively and might lead to (i) falsely concluding that a group of neurons do not encode a behavioral variable (when in fact they encode it collectively), (ii) incorrectly estimating the amount of information that is being encoded, and/or (iii) missing important mechanisms that contribute to encoding \citep{frost2021dynamic}. 

Alternatively, artificial neural networks (ANN) have been increasingly employed, either by (i) using a goal-driven neural network and using the embedding to compare and predict the population activity in different brain regions \citep{russo2018motor, mante2013context, jazayeri2021interpreting}, or (ii) modelling the activity of individual neurons as emanating from a feedforward or recurrent neural network architecture \citep{elsayed2016reorganization, rajan2016recurrent}. Whilst these methods can present powerful ways of inferring neural states and dynamics, some issues have been raised on their biological interpretability, even though recent work has addressed some of them, as we discuss in section \ref{S:ann}.

To address these shortcomings, a range of techniques which are commonly referred to under the umbrella term of ``neural manifold learning" (NML) have been employed. Some of these approaches simply make use of long-established general methods for dimensionality reduction (such as principal components analysis), whereas others have been developed specifically to study high-dimensional neural datasets. 

\begin{figure*}[!b]
\centering
\includegraphics[width=1\linewidth]{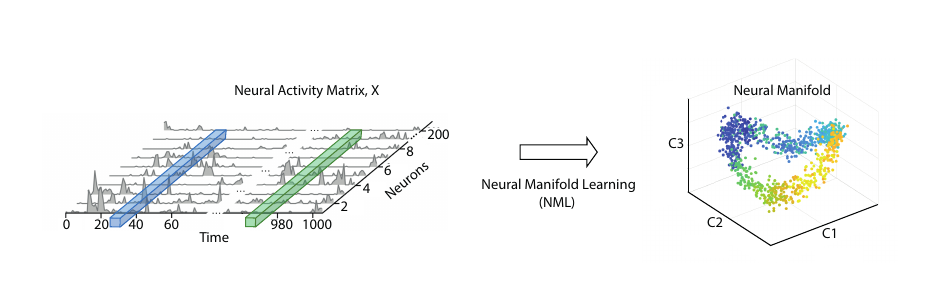}
\caption[dimred]{Schematic showing a typical example of how a manifold learning algorithm may reduce the dimensionality of a high-dimensional neural population time series to produce a more interpretable low-dimensional representation. A high-dimensional neural population activity matrix, $\mathbf{X}$, with $N$ neurons and $T$ time points, is projected into a lower-dimensional manifold space and the trajectory visualized in the space formed by the first 3 dimensions, $c1$, $c2$ and $c3$.}
\label{fig:schematic}
\end{figure*}

Mathematically speaking, a manifold is a topological space that locally resembles our usual Euclidean space. If we form a multivariate time-series by convolving the spike trains of a neural population with a smoothing filter, and consider the activity pattern across these time series at each time to occupy a point in a neural activity space, then over time the activity will excurse a subspace that has often been observed to appear like such a manifold. Characterizing the geometry of such structures may offer important insights into neural computation  \citep{chung2021neural}.  In practical terms, the ``neural manifold" is a low-dimensional subspace within the higher-dimensional space of neural activity which explains the majority of the variance of the neural dynamics (Figure \ref{fig:schematic}). Of course, real neural population dynamics are subject to noise, and in real experiments the topological subspace that can be excursed by the dynamics of neural activity can only be sampled, often sparsely. We must make several important comments here: firstly, that this characterisation of neural systems depends inherently upon a description of the system in which the state is continuous and determined by the instantaneous firing rates of each neuron (although those firing rates might be calculated from filters implementing shorter or longer time windows). And secondly, that what may be observed experimentally is typically a ``point cloud" - however these points do not themselves constitute a manifold; instead they are normally taken as indicative of the underlying topological space which they sample. And finally, it should be noted that the term ``neural manifold" is frequently used relatively loosely in neuroscience to refer to any kind of low-dimensional structure which may or may not meet the criteria of a mathematical manifold. In this survey article we will not dwell overly on this distinction, noting it but considering that utility is at this stage more important, and that it is likely that terminology will continue to evolve.

Neural manifold learning algorithms are algorithms for efficiently extracting a description of such a subspace from a sample of multivariate neural activity. Here we review how neural manifold learning can be employed to extract low-dimensional descriptions of the internal structure of ensemble activity patterns, and used to obtain insight into how interconnected populations of neurons represent and process information. Such techniques have been applied to neural population activity in a variety of different animals, brain regions and during distinct network states. In this review we compare a variety of neural manifold learning algorithms with several datasets from multi-electrode array electrophysiology and multi-photon calcium imaging experiments, to assess their relative merits in gaining insight into the recorded neural dynamics and the computations they may be underpinning.

We envisage that neural manifold learning may not only facilitate accurate quantitative characterizations of neural dynamics in healthy states, but also in complex nervous system disorders. Anatomical and functional brain networks may be constructed and analyzed from neuroimaging data in describing and predicting the clinical syndromes that result from neuropathology. NML can offer theoretical insight into progressive neurodegeneration, neuropsychological dysfunction, and potential anatomical targets for different therapeutic interventions. For example, investigating neural populations in the medial prefrontal cortex that are active during social situations while encoding social decisions may enable hypothesis testing for disorders such as Autistic Spectrum Disorder (ASD) or Schizophrenia \citep{irimia2018support, Kingsbury2019social}. 

In this review, we introduce NML as a methodology for neural population analysis and showcase its application to the analysis of different types of neural data with differing behavioral complexity both in healthy and disease model states. For selected algorithms, we visualize how they represent neural activity in lower-dimensional embeddings and evaluate them on their ability to discriminate between neural states and  reconstruct neural data. We aim to offer the reader the prospect of selecting with confidence the type of NML method that works best for a particular type of neural data, as well as an appreciation of how NML can be leveraged as a powerful tool in deciphering more precisely the basis of different cognitive impairments and brain disorders.

\section{Neural Manifold Learning}
\label{sec2}

\begin{figure*}[!b]
\centering
\includegraphics[width=1\linewidth]{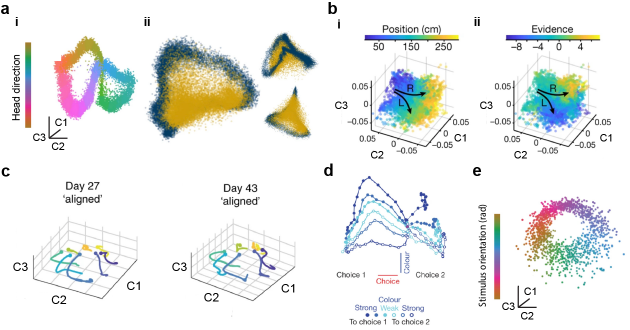}
\caption[dimred]{Neural manifolds across different species and brain regions. \textbf{(a)} Population activity in the mouse head-direction circuit \citep{Chaudhuri2019TheSleep}. (\textbf{i}) During waking, the network activity directly maps onto the head orientation of the animal. (\textbf{ii}) Comparison between population activity during waking (dark blue) and nREM sleep (mustard yellow); the latter does not follow the same one-dimensional waking dynamics. \textbf{(b)} Population activity in the mouse hippocampus during an evidence accumulation and decision making task in virtual reality \citep{nieh2021geometry}. Task-relevant variables such as (\textbf{i}) position and (\textbf{ii}) accumulated evidence are encoded in the manifold. \textbf{(c)} The motor cortical population activity underpinning a reaching task in monkeys is stable over days and years \citep{Gallego2020Long-termBehavior}. \textbf{(d)} Prefrontal cortical population activity in macaque monkeys during a visual discrimination task spans a low-dimensional space. Task-relevant variables such as the dots' direction and strength of motion, colour, and the monkey's choice are encoded in the manifold \citep{mante2013context}. \textbf{(e)} Population activity in the mouse primary visual cortex in response to gratings of different orientations, indicated by color \citep{stringer2021high}. The panel is adapted from \citet{jazayeri2021interpreting}.}
\label{fig:examples}
\end{figure*}

\noindent Neural manifold learning (NML) describes a subset of machine learning algorithms that take a high-dimensional neural activity matrix $\mathbf{X}$ comprised of the activity of $N$ neurons at $T$ time points and embed it into a lower-dimensional matrix $\mathbf{Y}$ while preserving some aspects of the information content of the original matrix $\mathbf{X}$ - e.g. mapping nearby points in the neural activity space $\mathbf{X}$ to nearby points in $\mathbf{Y}$ (see Figure \ref{fig:schematic}) \citep{cunningham2014dimensionality, Churchland2012NeuralReaching,meshulam2017collective,mante2013context,harvey2012choice,wu2017gaussian}. When projected into this lower-dimensional space, the set of neural activity patterns {\em observed} are typically constrained within a topological structure, or manifold, $\mathcal{Y}$, which might have a globally curved geometry but a locally linear one. For instance, if the reduced dimensionality embedding matrix $\mathbf{Y}$ is three-dimensional (3D) (often depicted on a two-dimensional (2D) page for convenience of illustration), the neural manifold $\mathcal{Y}$ might describe a closed surface within that 3D space. Another way of looking at this is that the data points in $\mathbf{X}$ lie on a lower-dimensional manifold that can be parameterised by a lower-dimensional coordinate system give by $\mathbf{Y}$, and that the task of the manifold learning algorithm is to find that coordinate system. This approach has found widespread recent use across neuroscientific studies (Figure~\ref{fig:examples}), including for understanding neural mechanisms during speech \citep{bouchard2013functional}, decision-making in prefrontal cortex \citep{mante2013context,harvey2012choice,briggman2005optical,stokes2013dynamic}, movement preparation and execution in the motor cortices \citep{Churchland2012NeuralReaching,kaufman2014cortical,yu2009gaussian,feulner2021neural,gallego2017neural} and spatial navigation systems \citep{Chaudhuri2019TheSleep, nieh2021geometry, rubin2019revealing, gardner2022toroidal}.

\subsection*{\textbf{Manifold learning algorithms}}
\label{S:methods}
\noindent There are several types of manifold learning algorithms that can generally be divided into linear and non-linear approaches. Although they have similar goals, they may differ in the way they transform the data and in the type of statistical structures or properties they capture and preserve. It is important to select the type of method  most suitable for the type of neural data being analyzed, as this may have significant impact on the resulting scientific interpretation. 

In this section, we describe some of the more common manifold learning methods in use in the neuroscience literature. To make a fair and informative comparison of each algorithm, we implemented them and applied them each to a number of different neural datasets, as described in Section \ref{S:manifoldcomparison}. In order to facilitate comparison of the assumptions made by the different algorithms, we have attempted to adopt (where possible) a common mathematical terminology throughout; this is summarised in Table~\ref{tab:notation}.

\begin{table}[t]
\centering
\begin{tabular}{l l l}
\hline
\textbf{Notation} & \textbf{Description}\\
\hline
$N$ & Number of neurons\\
$T$ & Number of time samples\\
$t$ & Sample timestamp\\
$\mathbf{X}$ & Population activity matrix\\
$\mathbf{x_t}$ & Network activity state at timestamp $t$\\
$k$ & Number of dimensions in the embedding\\
$\mathbf{Y}$ & Manifold embedding matrix\\
$\mathcal{Y}$ & Manifold - a topological structure within $\mathbf{Y}$\\
$\mathbf{D}$ & Dissimilarity matrix (MDS)\\
$\mathbf{G}$ & Graph of population activity\\
$\mathbf{\Lambda}$ & Graph Laplacian (LEM)\\
$\mathbf{\Delta}$ & Diagonal matrix (LEM)\\
$\mathbf{\Omega}$ & Adjacency matrix (LEM)\\
\hline
\end{tabular}
\newline
\caption{Summary of mathematical notation used throughout this manuscript.}
\label{tab:notation}
\end{table}

\subsection*{\textbf{\textit{Linear methods}}}
\label{S:linearmethods}
\noindent Linear manifold learning is accomplished by performing linear transformations of the observed variables that preserve certain optimality properties, which yield low-dimensional embeddings that are easy to interpret biologically. Some of the most common linear manifold learning techniques are discussed below.

\subsection{Principal Component analysis (PCA)}
\label{S:pca}
\noindent One of the most common linear manifold learning methods is Principal Component analysis (PCA) \citep{jolliffe2002principal,jackson2005user,ivosev2008dimensionality}. It reduces the dimensionality of large datasets while preserving as much variability (or statistical information) as possible. To obtain a neural manifold $\mathbf{Y}$, PCA performs an eigendecomposition on the covariance matrix of the neural activity matrix $\mathbf{X}$ in order to find uncorrelated latent variables that are constructed as linear combinations of the contributions of individual neurons, while successively maximising the variance. The computed eigenvectors, or principal components (PCs), represent the directions of the axes that capture the most variance, and the corresponding eigenvalues yield the amount of variance carried by each PC.

Although PCA has been used to great effect to reduce the dimensionality of high-dimensional neural data \citep{churchland2010cortical,mazor2005transient, gao2015simplicity,ahrens2012brain}, its caveat is that it captures all types of variance in the data, including spiking variability, which may obscure the interpretation of latent variables. For example, neurons with higher firing rates typically exhibit higher variance (spike count variance being proportional to mean spike count for cortical neurons \cite{tolhurst1983statistical}), and therefore may skew the orthogonal directions found by PCA by accounting mostly for highly active neurons. Additionally, cortical neurons respond with variable strength to repeated presentations of identical stimuli \citep{tolhurst1983statistical,shadlen1998variable,cohen2011measuring}. This variability is often shared among neurons, and such correlations in trial-to-trial responses can have a substantial effect on the amount of information encoded by a neuronal population. To minimise the effects of spiking variability, trial averaging or temporal smoothing of spike counts is usually done prior to performing PCA. However, this may not always be applicable to all analyses.

\subsection{Multidimensional Scaling (MDS)}
\label{S:mds}
\noindent Classical multidimensional scaling is another linear dimensionality reduction technique that aims to find a low-dimensional map of a number of objects (e.g. neural states) while preserving as much as possible pairwise distances between them in the high-dimensional space \citep{kruskal1978multidimensional,venna2006local,yin2008multidimensional,france2010two}. 

In the case of neural data, given an $N\times T$ activity matrix, $\mathbf{X}$, a $T\times T$ distance matrix, $\mathbf{D}$, is first obtained by measuring the distance between the population activity vectors for all $t$ using a dissimilarity metric \citep{krauss2018statistical}. Examples of such metrics include both the Euclidean distance and the cosine dissimilarity. We will employ the latter for the MDS embeddings shown throughout this paper. For a pair of vectors $\mathbf{a}$,$\mathbf{b}$ separated in angle by $\theta$, the cosine dissimilarity is
\begin{equation*}
    D_{\mathbf{a},\mathbf{b}}=\cos(\theta)=1-\frac{\mathbf{a}.\mathbf{b}}{\|\mathbf{a}\| \|\mathbf{b}\|} .
\end{equation*}
The dissimilarity matrix $\mathbf{D}$ is formed from the cosine dissimilarities between the neural activity patterns $\mathbf{c}_t$ at each time $t$ for all pairs of times measured.
A lower-dimensional mapping, $\mathbf{Y} \in \Re^{k \times T}$, where $k << N$, is then found by minimising a loss function, called the \textit{strain}, so that the mapped inter-point distances are as close as possible to the original distances in $\mathbf{D}$. From the eigen-decomposition of the distance matrix $\mathbf{D}$, the MDS components (i.e. dimensions) are revealed by the eigenvectors, which are a linear combination of the distances across the $T$ population activity vectors, while their respective eigenvalues report the amount of variance explained.

Multidimensional scaling has been used to study changes in neural population dynamics in both large-scale neural simulations \citep{phoka2012sensory} and in neural population recordings \citep{luczak2009spontaneous}. It has been applied for characterizing glomerular activity across the olfactory bulb in predicting odorant quality perceptions \citep{youngentob2006predicting}, integrative cerebral cortical mechanisms during viewing \citep{tzagarakis2009cerebral}, neuroplasticity in the processing of pitch dimensions \citep{chandrasekaran2007neuroplasticity}, emotional responses to music in patients with temporal lobe lesions \citep{dellacherie2011multidimensional}, and structural brain abnormalities associated with autism spectrum disorder \citep{irimia2018support}. 

\subsection*{\textbf{\textit{Non-linear methods}}}
\label{S:non-linearmethods}
\noindent The algorithms described above can only extract linear (or approximately so) structure. Non-linear techniques, on the other hand, aim to uncover the broader non-linear structure of the neural population activity matrix $\mathbf{X}$. These insights, can come at the expense of weaker biological interpretation, as the discovered manifold is not given by a linear combination of some observed variable (e.g. individual neurons). Non-linear algorithms often attempt to approximate the true topology of the manifold $\mathcal{Y}$ within the reduced dimensionality representation $\mathbf{Y}$ by finding population-wide variables that discard the relationship between faraway points (or neural states) on $\mathbf{X}$ and focus instead on conserving the distances between neighbors.

\subsection{Isomap}
\label{S:isomap}
\noindent One of the most commonly used non-linear manifold learning algorithms is Isomap \citep{tenenbaum2000global}. Non-linear techniques aim to uncover the broader non-linear structure of the neural manifold embedding of the neural population activity, $\mathbf{X}$, by approximating the true topology of the neural manifold, $\mathcal{Y}$. To do so, Isomap first embeds $\mathbf{X}$ into a weighted graph $\mathbf{G}$, whose nodes represent the activity of the neuronal population at a given time $\mathbf{x}_t$, and the edges between them represent links to network states ${\mathbf{x}_i , \mathbf{x}_j, ...}$ that are the most similar to $\mathbf{x}_t$, i.e., its neighbors. 

The $k$-nearest neighbors algorithm is usually used to estimate the neighbors for all network states $\mathbf{x}_1,..., \mathbf{x}_T$. These neighboring relations are then encoded in a weighted graph with edges $d_\mathbf{G}(i,j)$ between neighboring states $i,j$ that depends on the distance metric $d_\mathbf{X}$ used. $\mathbf{G}$ can then be used to approximate the true geodesic distances on the manifold $d_\mathcal{Y}(i,j)$ between any two points $i,j$ (i.e. network states $\mathbf{x}_i, \mathbf{x}_j$) by measuring their shortest path length $d_\mathbf{G}(i,j)$ in the weighted graph $\mathbf{G}$ using an algorithm such as Dijkstra's \citep{dijkstra1959note}. 

MDS is then applied to the matrix of shortest path lengths within the graph  $\mathbf{D}_\mathbf{G}=\{d_\mathbf{G}(i,j)\}$ to yield an embedding of the data in a $k$-dimensional Euclidean space $\mathbf{Y}$ that best preserves the manifold’s estimated geometry. The quality of the reconstructed manifold $\mathcal{Y}$ depends greatly on the size of the neighborhood search and the distance metric used to build $\mathbf{G}$. Isomap is a conservative approach that seems well suited to tackle the non-linearity inherent to neural dynamics and it has in fact been used in a variety of studies, even just for visualization purposes \citep{mimica2018efficient,Chaudhuri2019TheSleep,sun2019effective}.

\subsection{Locally Linear Embedding (LLE)}
\label{S:lle}
\noindent LLE is another non-linear manifold learning technique that attempts to preserve the local geometry of the high-dimensional data $\mathbf{X}$, by separately analyzing the neighborhood of each network state $\mathbf{x}_t$, assuming it to be locally linear even if the global manifold structure is non-linear \citep{roweis2000nonlinear}. The local neighborhood of each network state is estimated using the steps described in Section \ref{S:isomap} and is connected together to form a weighted graph $\mathbf{G}$. The location of any node $i$ in the graph corresponds to the network state $\mathbf{x}_i$ , which can then be described as a linear combination of the location of its neighboring nodes $\mathbf{x}_j, \mathbf{x}_k, ...$. These contributions are summarised by a set of weights, $\mathbf{W}$, which are optimised to minimise the reconstruction error between the high-dimensional network state $\mathbf{x}_i$ and the neighborhood linear estimation. The weight $w_{ij}$ summarizes the contribution of the $j^{th}$ network state to the reconstruction of the $i^{th}$ state. To map the high-dimensional dataset $\mathbf{X}$ onto a lower-dimensional one $\mathbf{Y}$, the same local geometry characterized by $\mathbf{W}$ is employed to represent the low-dimensional network state $\mathbf{y}_i$ as a function of its neighbors $\mathbf{y}_j, \mathbf{y}_k, ...$.

A significant extension of LLE was introduced that makes use of multiple linearly independent weight vectors for each neighborhood. This leads to a Modified LLE (MLLE) algorithm that is much more stable than the original \citep{zhang2007mlle}. Unlike Isomap, LLE preserves only the local geometry of the high-dimensional data $\mathbf{X}$, represented by the neighborhood relationship, so that short high-dimensional distances are mapped to short distances on the low-dimensional projection. In contrast, Isomap aims to preserve the geometry of the data at all scales, long distances included, possibly introducing distortions if the topology of the manifold is not estimated well. For neural data, though, Isomap has generally been preferred for its theoretical underpinnings and more intuitive approach.

\subsection{Laplacian Eigenmaps (LEM)}
\label{S:lem}
\noindent LEM, also referred to as Spectral Embedding, is another non-linear technique similar to LLE \citep{belkin2003laplacian}. The algorithm is geometrically motivated as it exploits the properties of the neighboring graph $\mathbf{G}$ generated from the high-dimensional data $\mathbf{X}$, as in Section \ref{S:isomap}, to obtain a lower-dimensional embedding $\mathbf{Y}$ that optimally preserves the neighborhood information of $\mathbf{X}$. 

In the graph $\mathbf{G}$, any two connected nodes (network states) $i$ and $j$ are connected by a binary edge using a neighborhood method as in \ref{S:isomap}, or by a weighted edge computed via a kernel parametrising the exponential relationship of the weights with respect to the distance between the nodes in the high-dimensional space $\mathbf{x}_i-\mathbf{x}_j$. The resulting graph edges form the adjacency matrix $\mathbf{\Omega}$ that is used, together with the diagonal matrix $\mathbf{\Delta}$ containing the degree of each node of $\mathbf{G}$, to obtain the graph Laplacian $\mathbf{\Lambda}=\mathbf{\Delta}-\mathbf{\Omega}$. The spectral decomposition of $\mathbf{\Lambda}$ reveals the structure and clusters on $\mathbf{G}$. The $k$ eigenvectors with the smallest non-zero eigenvalues of $\mathbf{\Lambda}$ are, in fact, the $k$ dimensions of the manifold embedding $\mathbf{Y}$. Similar to LLE, Laplacian eigenmaps preserve only the local geometry of the neural population activity $\mathbf{X}$ and are therefore more robust to the construction of $\mathbf{G}$. Indeed LEM has been successfully used to unveil behaviorally relevant neural dynamics \citep{rubin2019revealing, sun2019effective}.

\subsection{t-distributed Stochastic neighbor Embedding (t-SNE)}
\label{S:tsne}
t-SNE is another non-linear method that aims to match local distances in the high-dimensional space $\mathbf{X}$ to the low-dimensional embedding $\mathbf{Y}$. This is obtained by first constructing a probability distribution over pairs of high-dimensional points $\mathbf{x}_i, \mathbf{x}_j$ in such a way that nearby points are assigned a higher probability while dissimilar points are assigned a lower probability. Then t-SNE defines a similar probability distribution over the points $\mathbf{y}_i, \mathbf{y}_j$ in the low-dimensional space, and it minimizes the Kullback–Leibler divergence between the two distributions \citep{van2008visualizing}. The Euclidean distance is used in the original algorithm to evaluate the similarity between data points, but any appropriate metric can be employed as well. This method has been applied in a wide range of domains, from genomics to signal processing, including multiple neuroscientific settings \citep{dimitriadis2018t, panta2016tool}. It usually considers up to three embedding dimensions for visualization constraints, and for exploiting the Barnes-Hut approximation, which reduces the computational cost to $O(N\log N)$ from $O(N^2)$. \citep{tSNE_LaurensVDM}.

\subsection{Uniform Manifold Approximation and Projection (UMAP)}
\label{S:umap}
\noindent UMAP is a non-linear manifold learning technique that is constructed from a theoretical framework based on topological data analysis, Riemannian geometry and algebraic topology \citep{mcinnes2018umap} and it has also been used on neural data both as an NML method \citep{tombaz2020action} and  for broader dimensionality  reduction purposes \cite{lee2021non}. It builds upon the mathematical foundations of LEM, Isomap and other non-linear manifold learning techniques in that it uses a $k$ nearest neighbors weighted graphs representation of the data. By using manifold approximation and patching together local fuzzy simplicial set representations, a topological representation of the high-dimensional data is constructed. This layout of the data is then optimised in a low-dimensional space to minimize the cross-entropy between the two topological representations. Compared to t-SNE, UMAP is competitive in terms of visualization quality and arguably preserves more of the global dataset structure with superior run time performance. Furthermore, the algorithm is able to scale to significantly larger data set sizes than are feasible for t-SNE.

UMAP and t-SNE have recently being employed for visualizing high-dimensional genomics data and some distortion issues have been raised \citep{Chari2021TheGenomics}. Although this problem is particularly apparent with t-SNE, which tends to completely disregard the global structure of the data to find clusters, any reduction to too few dimensions with respect to the original high-dimensional space will inherently distort the topology of some of the data, as indicated by the Johnson-Lindenstrauss Lemma \citep{Johnson1984ExtensionsSpace}. These criticisms have been made with particular reference to genomics datasets, which are intrinsically higher-dimensional than the neural datasets which manifold learning has been applied to.

\subsection{Probabilistic latent variable models}
\label{S:gpfa}
\noindent Another type of NML algorithm uses probabilistic latent variable models that construct a generative model for the data in terms of mapping a low-dimensional manifold or latent space to neural responses. This type of algorithm utilizes a probabilistic framework that performs temporal smoothing and dimensionality reduction simultaneously, allowing joint optimisation of the degree of smoothing and the relationship between the original high-dimensional data and the resulting low-dimensional neural trajectory \citep{yu2009gaussian}. A good example is Gaussian Process Factory analysis (GPFA) that uses Gaussian processes and an additional explicit noise model to account for the different independent noise variances of different neurons (i.e., spiking variability). It is a set of factor analyzers that are linked together in the low-dimensional state space by a Gaussian process prior \citep{rasmussen2006gaussian}, which allows for the specification of a correlation structure across the low-dimensional states at different time points. In cases where the neural time courses are believed to be similar across different trials, smooth firing rate profiles may be obtained by averaging across a small number of trials \citep{mazor2005transient,stopfer2003intensity,brown2005encoding,broome2006encoding,levi2005role,nicolelis1995sensorimotor}, or by applying more advanced statistical methods for estimating firing rate profiles from single spike trains \citep{dimatteo2001bayesian,cunningham2007inferring} 
Similarly, Manifold Inference from Neural Dynamics (MIND) is a recently developed NML algorithm that aims to characterize the neural population activity as a trajectory on a non-linear manifold, defined by possible network states and temporal dynamics between them \citep{low2018probing, nieh2021geometry}.

\subsection{NML algorithms for trial-structured datasets}
\label{S:tNMLS}
Importantly, model selection from a computed manifold can be greatly affected by the signal-to-noise ratio (SNR) of the initial input neural data. In many cases this has been overcome by using the square root transformation of spiking data and convolving it with a Gaussian filter to yield a smoothed instantaneous firing rate \citep{yu2009gaussian}. In addition, multiple dimensionality reduction steps have also been used to enable more interpretable visualizations (LEM on LEM \citep{rubin2019revealing}, UMAP on PCA \citep{gardner2022toroidal}).
 Furthermore, the NML algorithms described and visualized up until now have been general use case algorithms, used to infer neural correlates from the data.  However, in experiments where specific behaviors or decisions are time-locked and run across multiple trials, some of the NMLs described above have been augmented and optimised.  These include, demixed Principal Component analysis (dPCA) \citep{kobak2016dpca}, Tensor Component analysis (TCA) \citep{cohen2010neuronal,niell2010modulation,peters2014emergence, driscoll2017dynamic}, Cross-Validated PCA (cvPCA) \citep{stringer2019high} and model-based Targeted Dimensionality Reduction (mTDR) \citep{Aoi2018mTDR}.  These NML algorithms exploit the trial nature of an experiment to discriminate signal from trial-to-trial variability or noise, enabling the experimenter to identify the principal components that maximally correspond to a stimulus or action.

\subsection{ANN-based NML algorithms}
\label{S:ann}
\noindent Artificial neural networks (ANN) can also be employed for manifold learning as they have the potential to extract complex non-linear structure in high-dimensional data. Auto-encoders exemplify this approach as they are designed to find an optimal encoding between a high-dimensional input and a low-dimensional representation stored in their "bottleneck" code layer, which preserves the information necessary to then reconstruct the original input from it. In this sense, auto-encoders can be thought of as a non-linear extension to PCA, where each node in the code layer is comparable to a PC. Delving deeper, the field of deep generative models such as variational auto-encoders (VAE) promise great potential at extracting low-dimensional structure in varied high-dimensional data, by constructing a stochastic model of the low-dimensional dynamics underlying the neural activity. Such methods have shown great results when inferring the neural activity trial-by-trial fluctuations, but some have raised the issue that the low-dimensional structure extracted from these models are often highly entangled and therefore, difficult to interpret \citep{pandarinath2018inferring}. To address these issues, VAEs that make use of external labels, such as behavioral variables or the passage of time, have been designed \citep{zhou2020learning}. Lastly, addressing some of the shortcomings of VAEs such as interpretability, identifiability and generalizability, Consistent EmBeddings of high-dimensional Recordings using Auxiliary variables, or CEBRA, was developed. CEBRA uses an innovative approach that employs contrastive learning instead of a generative model to extract embeddings, this enables it to cope with strong data distribution shifts to yield consistent embeddings between experimental sessions, subjects and recording modalities \citep{schneider2022learnable}.

\section{\textbf{Manifold learning for the analysis of large-scale neural datasets}}
\label{S:manifoldcomparison}
\noindent To demonstrate how neural manifold learning can be used in the analysis of large-scale neural datasets, we apply a number of linear and non-linear NML algorithms (PCA, MDS, LEM, LLE, Isomap, t-SNE and UMAP) to several datasets. The datasets were chosen to cover a number of different brain regions and a range of behavioral complexity. They consist of (i) two-photon calcium imaging of hippocampal subfield CA1 in a mouse running along a circular track (Section \ref{S:hippocampus}), taken from \cite{GoPlacecells}; (ii) multi-electrode array extracellular electrophysiological recordings from the motor cortex of a macaque performing a radial arm goal-directed reach task (Section \ref{S:monkey}) from \cite{yu2007mixture}; and (iii) single-photon "mini-scope" calcium imaging data recorded from the prefrontal cortex of a mouse under conditions where multiple task-relevant behavioral variables were monitored (Section \ref{S:acc}), from \cite{rubin2019revealing}. Lastly, we illustrate how manifold learning can be employed to characterize brain dynamics in a disease state such as Alzheimer's disease by applying these techniques to data simulated to reproduce basic aspects of dataset (i), augmented to incorporate pathology (Section \ref{S:AD}).

\subsection*{Decoding from neural manifolds}
\label{S:decode}
\noindent To compare NML algorithms we evaluated the resulting manifolds according to behavioral decoding (reconstruction or classification) performance, ability to encode the high-dimensional activity (i.e reconstruction score) and intrinsic dimensionality. These quantifications make up a minor subset of the many manifold parameterization methods, of which we describe more section \ref{S:parameterization}). For hippocampal CA1 manifolds obtained by any of the NML methods, we computed decoding accuracy for the behavioral variable(s) as a function of the number of manifold embedding dimensions using an Optimal Linear Estimator (OLE) \citep{warland1997decoding}. This allows assessment of the number of dimensions necessary to encode the behavioral variable. We used a 10-fold cross-validation approach, i.e., training the decoder on 90\% of the data and testing it on the remaining 10\%. Decoding performance is calculated as the Pearson correlation coefficient between the actual and reconstructed behavioral variable, i.e. the mouse position, for the test data. To assess neural manifold information provided about animal behavior in the other two datasets, we built a logistic regression classifier \citep{hosmer1989best}; we evaluate its performance using the F1 score as a function of the number of manifold embedding dimensions used. The F1 score is defined as the weighted average of the precision (i.e., percentage of the results which are relevant) and recall (i.e., percentage of the total relevant results correctly classified by the algorithm), and ranges between 0 (worst) and 1 (best performance) \citep{blair1979information}.

\subsection*{Reconstruction of neural activity from a low-dimensional embedding}
\label{S:reconstruct}
\noindent Another way to evaluate the degree of fidelity of the manifold embedding is to attempt to reconstruct the high-dimensional neural data from the low-dimensional embedding. This tells us how much has been lost in the process of dimensionality reduction. To obtain such a reverse mapping, we employed the non-parametric regression method originally introduced for LLE \citep{low2018probing, nieh2021geometry}. We then obtained the reconstruction similarity by computing the Pearson correlation coefficient between the reconstructed and the original neural activity. To perform an element-wise comparison, the $N\times T$ neural activity matrices were concatenated column-wise into a single vector and the correlation coefficient calculated. To obtain the neural activity reconstruction score, we employed a 10-fold cross-validation strategy. Using 90\% of the data from each session to learn the reverse mapping, the reconstruction was then evaluated on the remaining 10\% the data. The final score was then obtained by averaging across folds.

\subsection*{Intrinsic manifold dimensionality}
\label{S:dimensionality}
\noindent Estimating the number of required dimensions of the underlying manifold is a crucial part of manifold learning \citep{altan2021estimating,cunningham2014dimensionality}, as it helps one to acquire a conceptual idea of how complex the neuronal activity inside the manifold is. The intrinsic manifold dimensionality accounts for the number of independent (latent) variables necessary to describe the neural activity without suffering significant information loss \citep{jazayeri2021interpreting}. However, it is difficult to estimate the dimensionality of neural manifolds, especially in the realistic condition of a noisy, non-linear embedding. A recent review provides a detailed evaluation of several dimensionality estimation algorithms when applied to high-dimensional neural data \citep{altan2021estimating}.

We acknowledge that any measure of dimensionality is strongly influenced by the timescale of the neural activity and by the size of the population recorded \citep{humphries2020strong}, in this review we use the intrinsic dimensionality measure to compare the topologies captured by the different NML methods. To infer the manifold intrinsic dimensionality we employ a method related to the correlation dimension of a dataset \citep{grassberger1983characterization}. After applying NML to the original high-dimensional neural activity, for any given point (i.e., neural state) in the low-dimensional space, the number of neighboring neural states within a surrounding sphere as a function of the sphere’s radius was calculated. The slope of the number of neighboring data points within a given radius on a log–log scale equals the exponent $k$ in the power law $N(r) = cr^k$, where $N$ is the number of neighbors and $r$ is the sphere radius. $k$ then provides an estimate of the intrinsic dimensionality of the neural activity. We fit the power law in the range between $\sim$10 to 5000 neighbors or neural states, aiming to capture the relevant temporal scale for each task and related manifold \citep{rubin2019revealing, nieh2021geometry}. Note, in fact, that we have selected a particular range for each dataset, also depending on the number of time samples $T$ available (details in respective figure captions). 

\subsection{Hippocampal neural manifolds}
\label{S:hippocampus}
\noindent The hippocampus is well known to be involved in memory formation and spatio-contextual representations \citep{scoville1957loss, o1978hippocampal, morris2006elements}. NML has been recently applied to hippocampal neural activity by several authors, suggesting that the rodent hippocampal activity encodes various contextual, task-relevant variables, displaying more complex information processing than spatial tuning alone \citep{rubin2019revealing,nieh2021geometry}. Here, we re-analyze published data from a dataset comprising two-photon calcium imaging of hippocampal CA1 place cells, to which previously only MDS had been applied \cite{GoPlacecells}, in order to compare manifolds extracted by different algorithms. We examine how different NML methods characterize the dynamics of hippocampal CA1 neurons along trajectories in low-dimensional manifolds as they coordinate during the retrieval of spatial memories.

\begin{figure*}[ht!]
\begin{center}
    \includegraphics[width=1\linewidth]
{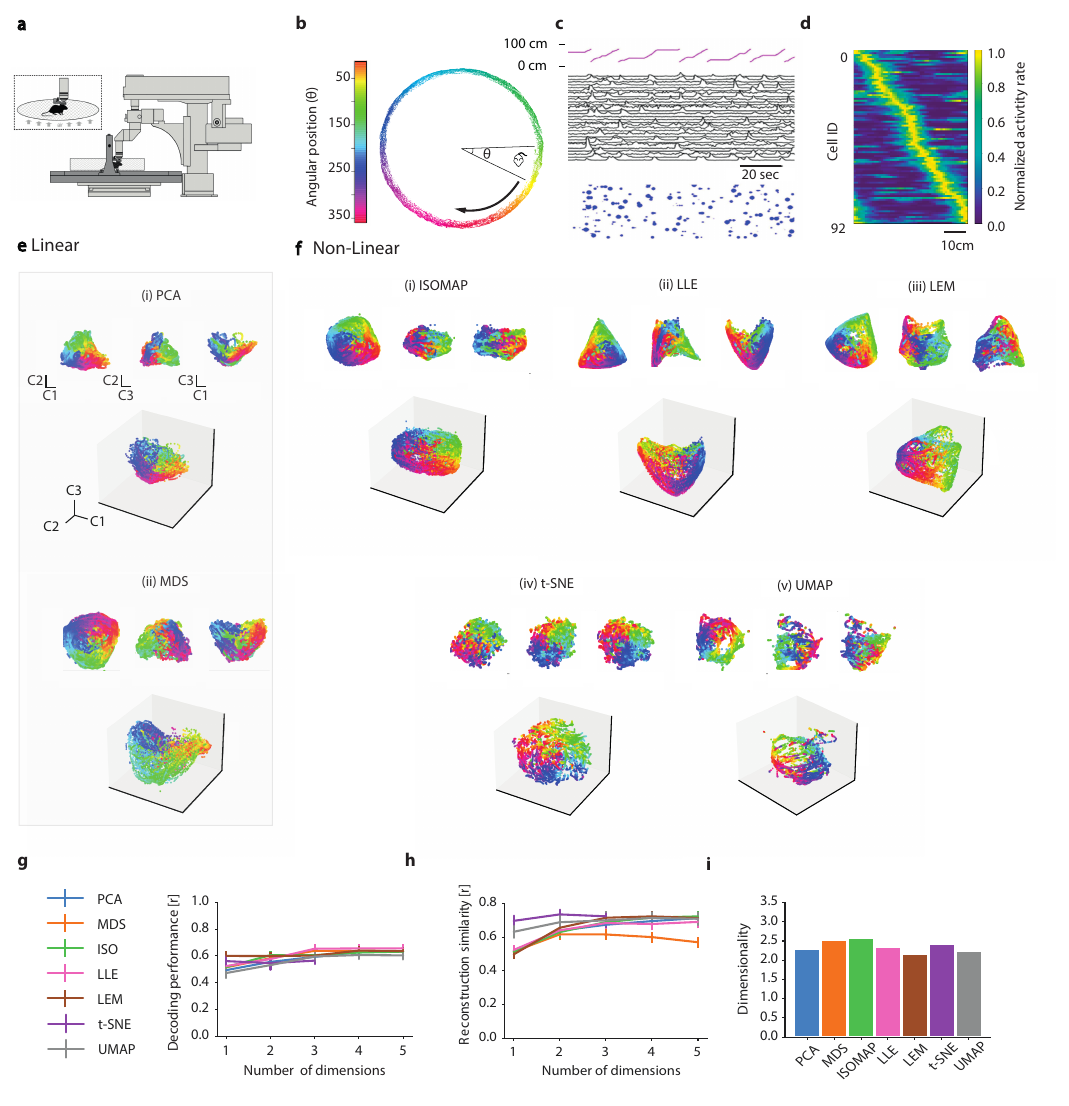}
\end{center}
\caption[Mouse hippocampal CA1 manifolds during spatial memory retrieval]{\textbf{Mouse hippocampal CA1 manifolds during spatial memory retrieval.} Unless otherwise stated,  panels adapted with permission from \citet{GoPlacecells}. \textbf{(a)} Schematic of experimental setup: head-fixed mouse navigates a floating circular track under a two-photon microscope. Inset: Close-up view of head-fixed mouse on floating track. \textbf{(b)} Spatial trajectories of the mouse, with $\theta$ (color-coded) denoting position on the track. \textbf{(c)} Top: Position along the track (cm), Middle: Ca\textsuperscript{2+} transients for 30 (of 225) randomly selected cells, Bottom: Rastergram showing events detected from the above calcium transients. Blue dots indicate time of Ca$^{2+}$ transient onset, with dot area showing relative event amplitude (normalized per cell). \textbf{(d)} Normalized neural activity rate map for 92 place cells, sorted by spatial tuning around the track.
\textbf{(e - f)} Linear vs non-linear manifold embeddings for all 225 cells (normalized).  In each case the first three dimensions are visualized. Insets for each: projections on pairs of components C1 and C2 (upper left), C2 and C3 (upper middle), C1 and C3 (upper right). \textbf{(e)} Linear manifold embeddings  (i) PCA,  (ii) MDS, \textbf{(f)} non-linear manifold embeddings (i) Isomap, (ii) LLE, (iii) LEM, (iv) t-SNE, and (v) UMAP. (ii-v were not reproduced from \cite{GoPlacecells}. \textbf{(g-i)} Manifold evaluation metrics: (see: \textbf{(g)} Decoding performance (as used in \citet{GoPlacecells}), \textbf{(h)} Neural activity reconstruction score, \textbf{(i)} Intrinsic manifold dimensionality.}
\label{fig:CA1}
\end{figure*}

The data was recorded from a head-fixed mouse navigating a circular track in a chamber floating on an air-pressurized table under a two-photon microscope (Figure \ref{fig:CA1}a). The mouse position (Figure \ref{fig:CA1}b) was simultaneously tracked using a magnetic tracker. The activity of 30 of 225 hippocampal CA1 cells recorded in the shown session is depicted in Figure \ref{fig:CA1}c. Of the 225 cells, 92 were classified as place cells by \citet{GoPlacecells}, and their normalized activity rate map, sorted according to place preference, is shown in Figure \ref{fig:CA1}d. Employing the activity of all 225 cells (both place and non-place selective), both linear (Figure \ref{fig:CA1}e) and non-linear (Figure \ref{fig:CA1}f) NML methods, revealed a cyclic pattern of transitions between network states corresponding to different locations along the circular track. The manifold formed by the dynamics of neural activity as the mice explored the full track forms a complete representation of the 2D structure of the track. We compared the algorithms in terms of decoding performance (Figure \ref{fig:CA1}g), neural activity reconstruction score (Figure \ref{fig:CA1}h) and intrinsic manifold dimensionality (Figure \ref{fig:CA1}i). All algorithms performed similarly in terms of the metrics considered, yielding almost the best possible decoding performance with just one manifold dimension and the best possible reconstruction similarity with two dimensions. Moreover, all manifolds were found to have a similar intrinsic dimensionality of around 2.

In this example, the behavioral complexity is approximately one dimensional (i.e., the mouse running in a single direction along a circular track can be mapped onto the single circular variable $\theta$) and all NML methods produce embeddings which allow high decoding performance, with each algorithm already reaching near-maximum performance after incorporating only the first manifold dimension. This suggests that if the behavioral complexity is low and its information is broadly encoded within the neural population, any NML algorithm will yield broadly similar results. However, as we will see, this does not necessarily hold when complexity of the behavioral task is increased. In terms of the ability to capture the neural activity variance, the neural activity reconstruction score suggests that the highly non-linear tSNE and UMAP algorithms yield more informative low-dimensional embeddings.

\subsection{Motor cortical neural manifolds}
\label{S:monkey}
\noindent NML and dimensionality reduction techniques have also been applied to neural activity within motor and pre-motor cortical areas, in particular to suggest that the high variability observed in the single-neuron responses is disregarded when population dynamics are taken into account \citep{Santhanam2009FA, churchland2007temporal, Churchland2012NeuralReaching, sussillo2015neural, gallego2017neural}. This approach has been the foundation of the "computation through dynamics" framework, which aims to characterize the neural population dynamics as trajectories along manifolds \citep{vyas2020computation}. Rotational-like dynamics of motor cortex neural activity have been observed both in human and non-human primates \citep{Churchland2012NeuralReaching, pandarinath2015neural}, with stability over long periods of time \citep{Gallego2020Long-termBehavior}. Importantly, by furthering the understanding of neural population dynamics and variability, crucial steps can be made in improving performance of prosthetic devices that can be used to further enable those with nervous-system disease or injury in day-to-day tasks \citep{Santhanam2009FA}.

\begin{figure*}[ht!]
\begin{center}
    \includegraphics[width=0.95\linewidth]{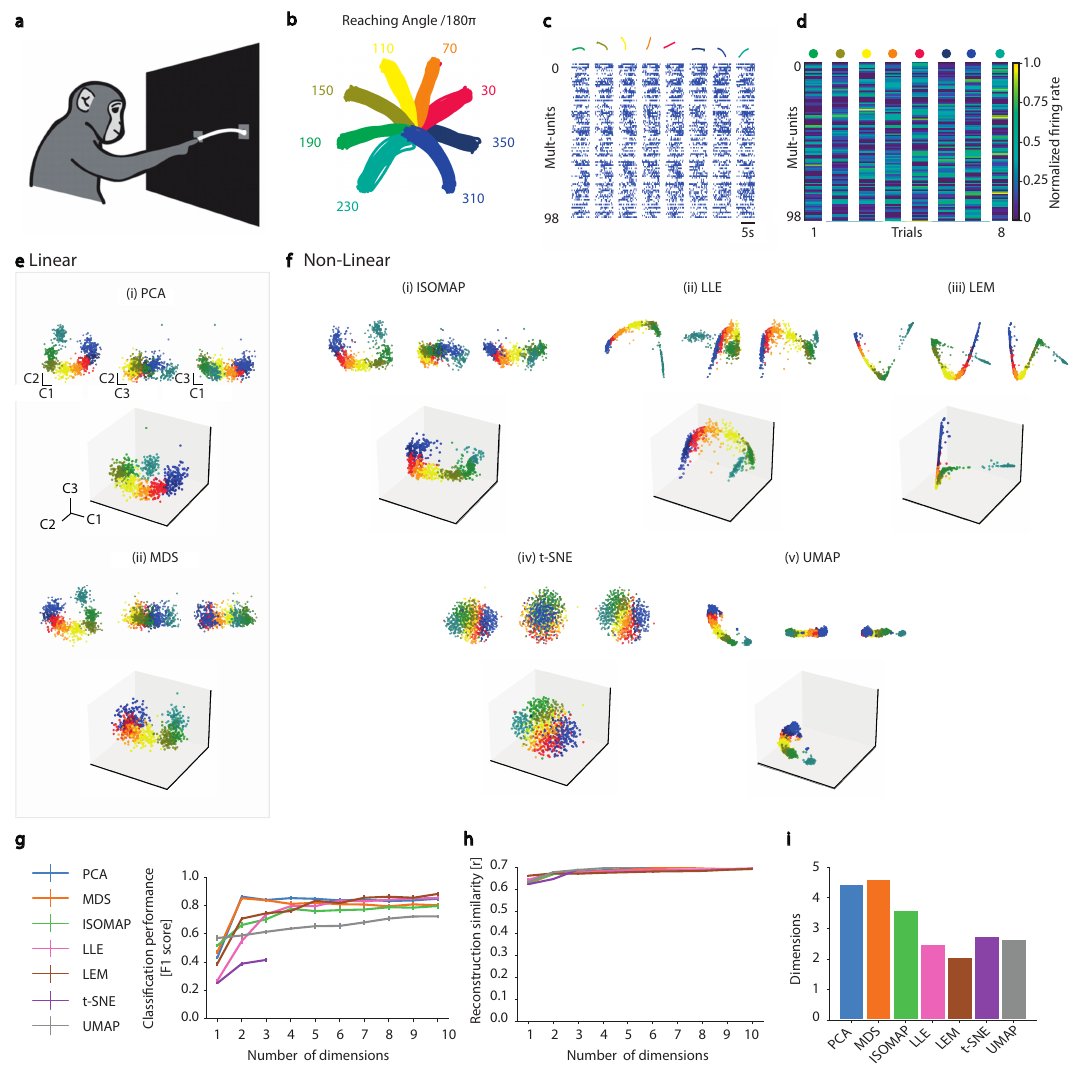}
\end{center}
  \caption[Manifold analysis of motor cortical dynamics during a reaching task.]{\textbf{Manifold analysis of motor cortical dynamics during a reaching task.} \textbf{(a)} Monkey reaching task experimental setup. \textbf{(b)} Arm trajectories and target reaching angles - schematic and coloured trajectories adapted from \cite{yu2007mixture}. \textbf{(c-d)} Neural data processing steps (adapted from \citet{Santhanam2009FA}) \textbf{(c)} Top: X,Y position of monkey arm over eight individual trials coloured according to (b). Bottom: Corresponding raster plot of spike trains recorded simultaneously from 98 neural units, sampled every 1 ms. Units include both well isolated single-neuron units (30\% of all units), as well as multi-neuron units. \textbf{(d)} Population activity vector yielding the instantaneous firing rates of the same neural units in (c) for the 100 ms before the monkey reached out its arm (separated and coloured by trial type/arm direction as in c). \textbf{(e - f)} Linear vs non-linear manifold embeddings (normalized, only first three dimensions visualized).  Insets: projections on pairs of components, C1 and C2 (upper left), C2 and C3 (upper middle), C1 and C3 (upper right). \textbf{(e)} Linear manifold embeddings (i) PCA, (ii) MDS. \textbf{(f)} non-linear manifold embeddings (i) Isomap, (ii) LLE, (iii) LEM, (iv) t-SNE, and (v) UMAP \textbf{(g-i)} Manifold evaluation metrics \textbf{(g)} Decoding performance [F1 score], \textbf{(h)} Neural activty reconstruction score [r], \textbf{(i)} Intrinsic manifold dimensionality.}
\label{fig:motor}
\end{figure*}

We applied NML to data from \citet{yu2007mixture} and \cite{chestek2007single} recorded in the caudal premotor cortex (PMd) and rostral primary motor cortex (M1) of rhesus macaques during a radial arm goal-directed reaching task (Figure \ref{fig:motor}a). Neural data was collected for 800 successful repeats, with 100 trials for each of the 8 reaching directions (Figure \ref{fig:motor}b). We used NML to analyze the neural activity during the 100 ms time window before the onset of the reaching movement and investigated its tuning with respect to the reaching direction \citep{Santhanam2009FA}. The manifold embeddings obtained using different NML methods, both linear (Figure \ref{fig:motor}e) and non-linear (Figure \ref{fig:motor}f), revealed different types of structures with data points clustering according to the monkey's target endpoint. All the NML algorithms tested revealed lower-dimensional structures that discriminate between each of the behavioral states along a single dimension. We used the pre-movement neural manifold to classify the behavior into one of eight different goal directions. The two linear NML algorithms yielded the most behaviorally informative embedding, requiring only two dimensions to achieve best performance (Figure \ref{fig:motor}g). All algorithms performed equally in terms of neural activity reconstruction similarity, with only one dimension being necessary to reconstruct the original neural activity patterns (Figure \ref{fig:motor}h). The intrinsic dimensionality of the two linear embeddings, on the other hand, was the highest (Figure \ref{fig:motor}i). Non-linear NML algorithms extracted lower-dimensional embedding at the cost of encoding less behavioral information.

\subsection{Prefrontal cortical neural manifolds} 
\label{S:acc}

\begin{figure*}[ht!]
\begin{center}
    \includegraphics[width=0.95\linewidth]{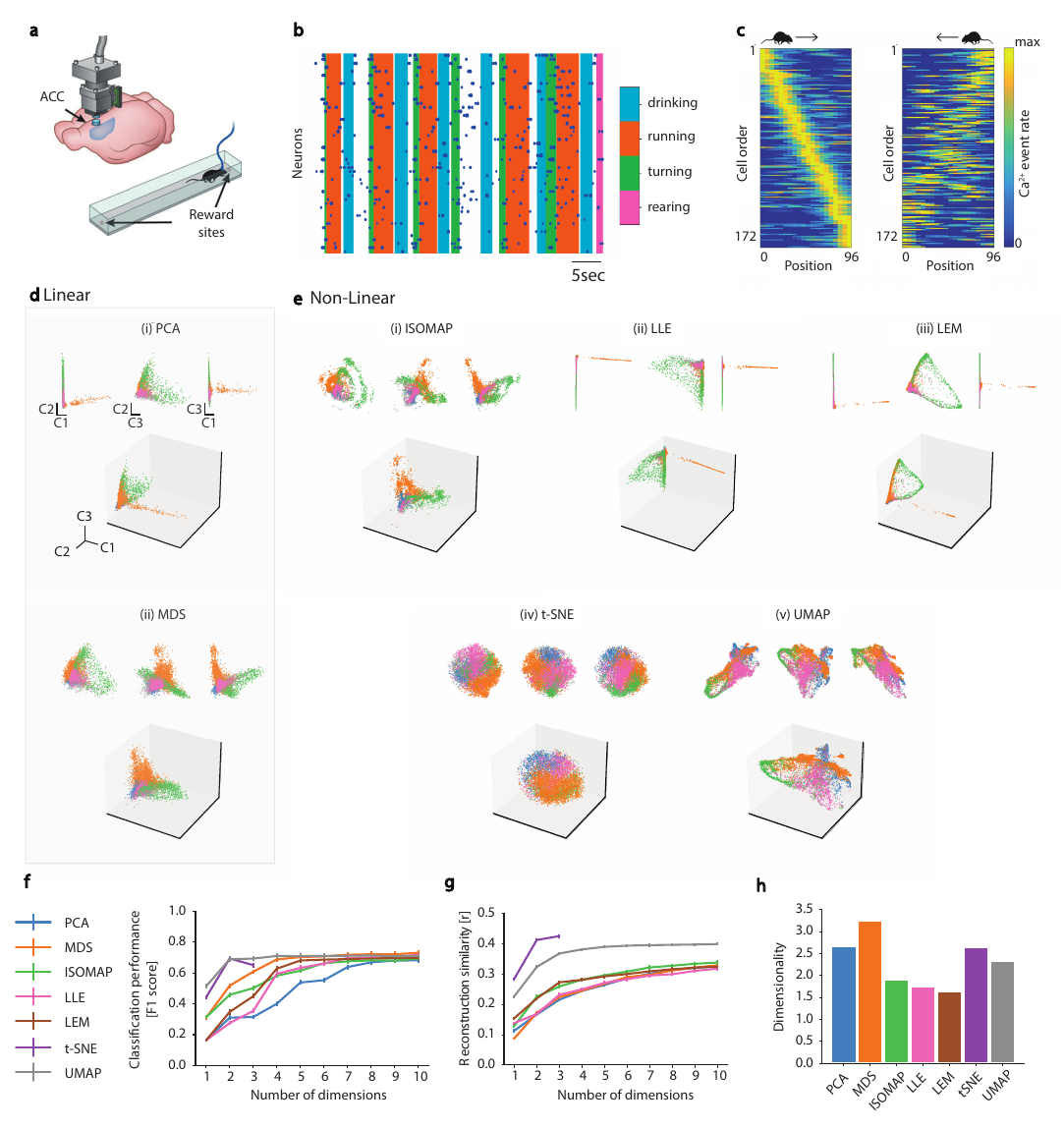}
\end{center}
\caption[Manifolds of ACC dynamics during multiple behaviors]{\textbf{Manifold analysis of prefrontal cortical dynamics during behavior.} \textbf{(a-c)} Schematic of experimental setup from \citet{rubin2019revealing}. The mouse navigates a linear track with rewards at both ends. \textbf{(b)} Raster plot of the neural activity during multiple behaviors (adapted from \citet{rubin2019revealing}): (1) drinking, (2) running, (3) turning, and (4) rearing. \textbf{(c)} Calcium event rate as the mouse runs to the right (left panel) and to the left (right panel) of the linear track (from \citet{rubin2019revealing}). \textbf{(d-e)} Linear vs non-linear manifold embeddings (normalized).  In each case the first three dimensions are visualized: Insets for each: projections on pairs of components: C1 and C2 (upper left), C2 and C3 (upper middle), C1 and C3 (upper right).  \textbf{(d)} Linear manifold embeddings  (i) PCA,  (ii) MDS, \textbf{(e)} non-linear manifold embeddings (i) Isomap, (ii) LLE, (iii) LEM, (iv) t-SNE, and (v) UMAP. \textbf{(f-h)} Manifold evaluation metrics \textbf{(f)} Decoding performance [r], \textbf{(g)} Reconstruction score [r], \textbf{(h)} Intrinsic manifold dimensionality.}
\label{fig:prefrontal}
\end{figure*}

\noindent Increasingly, NML has been used to analyze neural circuits implicated in decision-making such as the prefrontal cortex (PFC) \citep{Kingsbury2019social, mante2013context} and the anterior cingulate cortex (ACC) \citep{rubin2019revealing}. With these multi-function brain regions, neurons are widely known for their mixed selectivity, often capturing information relating to both stimuli features and decisions \citep{kobak2016dpca, rigotti2013importance, fusi2016neurons}. This moves away from the notion of highly-tuned single cells and gives rise to a dynamical, population-driven code \citep{Pouget2000} ideally suited to NML methods. This was nicely demonstrated by \citet{rubin2019revealing}, who visualized and quantified how NML (in their case, LEM) could be used to discriminate neural states arising from different brain regions (ACC and hippocampal area CA1) during identical tasks. The neural activity was recorded in freely-behaving mice exploring a linear track and performing various behaviors such as drinking, running, turning, and rearing (Figure \ref{fig:prefrontal}a-c). Building on their findings, we employed various alternative NML methods, both linear (Figure \ref{fig:prefrontal}d) and non-linear (Figure \ref{fig:prefrontal}e), that revealed different neural data structures, exhibiting clustering according to the animals' behavior, with PCA and LLE producing the least clustered visualizations. Evaluation of NML behavior classification performance (Figure \ref{fig:prefrontal}f), reconstruction similarity (Figure \ref{fig:prefrontal}g) and manifold intrinsic dimensionality (Figure \ref{fig:prefrontal}i) revealed that although there was high variability in performance, non-linear algorithms generally out-performed the linear ones.  Notably, two non-linear NML algorithms t-SNE and UMAP performed the best in terms of behavioral classification and ability to reconstruct the high-dimensional neural activity from the manifold embedding, with an intrinsic dimensionality between 2 and 3, inferring that behavior is the predominant type of information encoded by the ACC network during this task.

\subsection{Analysis of neural manifolds in neurological disorders}
\label{S:AD}
\noindent NML potentially provides a valuable tool for understanding the biology of different brain disorders. As an example, NML can be used to characterize changes in neural manifolds for spatial memory during the progression of Alzheimer's disease (AD). In AD, the hippocampus and connected cortical structures are among the first areas to show pathophysiology. Hippocampal-dependent cognitive processes such as episodic memory are particularly and prominently affected at the behavioral level. However, it is not yet understood how the pathological markers of AD, such as amyloid-beta plaques and neurofibrillary tangles, lead to specific disruptions in the network-level information processing functions underpinning memory function. While single-cell properties are obviously affected, it is believed that network properties relating to the coordination of information throughout brain circuits involved in memory may be particularly at risk \citep{palop2016network}. Neural manifold learning analysis methods may thus play a useful or even crucial role in disentangling the effects of these network alterations.

\begin{figure*}[ht!]
\begin{center}
    \includegraphics[width=\textwidth, height=18cm]{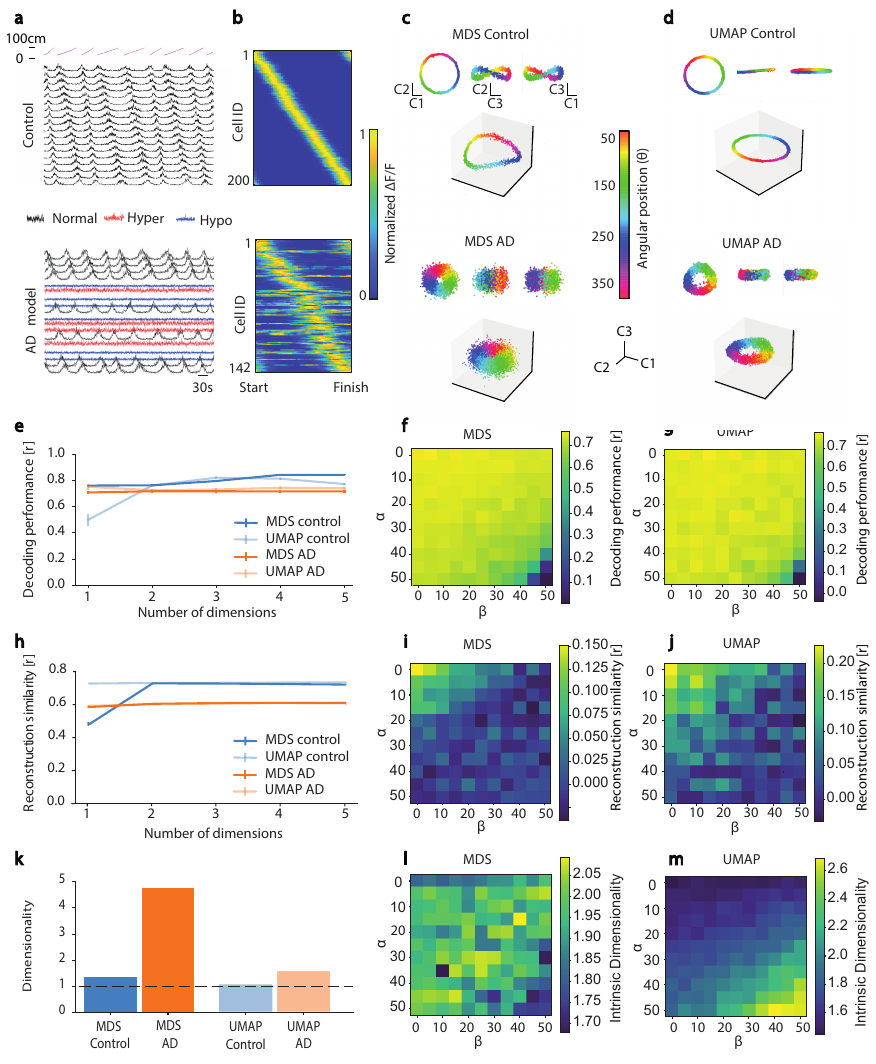}
\end{center}
\caption[Neural manifolds as a means of analyzing neurodegenerative disease]{\textbf{Changes in neural manifolds during neurodegenerative disease states}. \textbf{(a)} Example simulated calcium traces for a mouse traversing a circular track (simulation of real experiment in Figure~\ref{fig:CA1}). Top: 20 of 200 ``normal" CA1 place cells (magenta traces on top show mouse trajectory). Bottom: 20 of 200 AD-affected place cells that fall into three categories: Normal (black), Hyperactive (red) and Hypoactive (blue). \textbf{(b)} Normalized activity rate maps for 200 healthy and 142 (out of 200) AD hippocampal CA1 place and hyperactive cells (respectively) sorted by location of maximal activity. \textbf{(c-d)} 3D projection in the manifold space using (c) MDS and (d) UMAP for both healthy control and AD simulated models. Inset: 2D projections onto pairs of dimensions.  \textbf{(e-m)} Evaluation metrics for MDS and UMAP, Control vs AD and shown as a function of varying percentages of normal, hyperactive ($\alpha$) and hypoactive ($\beta$) cells. \textbf{(e-g)} Decoding performance [r]. \textbf{(h-j)} Reconstruction similarity [r]. \textbf{(k-m)} Intrinsic dimensionality.}
\label{fig:disease}
\end{figure*}

The formation of extracellular amyloid plaques causes aberrant excitability in surrounding neurons \citep{busche2008clusters}: cells close to amyloid plaques tend to become hyperactive, whereas those further away from the plaques tend to become hypoactive (or relatively silent). This aberrant neuronal excitability could potentially have a substantial effect on the neural manifold, and conversely, the neural manifold may provide a way to assess the overall impact of network abnormalities. To demonstrate this, we simulated neural data for the  \citep{GoPlacecells} circular track experiment described in \ref{fig:CA1}) using parameters outlined by \citep{busche2008clusters}, i.e., the percentages of hyperactive ($\alpha$) and hyperactive cells ($\beta$) in a given FOV, while conforming to previous studies that observed more than 20\% of neurons becoming hyperactive \citep{busche2008clusters,busche2015neuronal} in transgenic AD mice. This yielded a population of hippocampal CA1 cells (with 50$\%$ normal cells, 21$\%$ hyperactive cells and 29$\%$ hypoactive cells). Hypersynchrony, which has also been observed in AD \citep{palop2016network}, was not simulated for the purposes of this exercise, to maintain simplicity. This disease model mimics the neural activity of hippocampal CA1 cells of an AD mouse (e.g. a mouse over-expressing amyloid precursor protein, \citep{oakley2006intraneuronal}) running on a circular track; we compare it with a simulated control model representing a healthy wild-type litter-mate (Figure \ref{fig:disease}a,b). Using MDS and UMAP as exemplars (Figure \ref{fig:disease}c and \ref{fig:disease}d, respectively), NML shows how the aberrant excitability of neurons distorts the neural manifolds for spatial memory in AD (bottom panel) with respect to the healthy control model (top panel). While the topological structure remains relatively intact, the boundaries of the set become fuzzier, akin to adding noise to the system. To further evaluate this, we examined manifold parameterization measures including decoding performance, neural reconstruction similarity and manifold intrinsic dimensionality for both NML approaches (Figure \ref{fig:disease}e-m). In both cases, the AD model's manifold embeddings display worse performance than the control model, although UMAP was able to recover a more behaviorally informative and lower-dimensional embedding than MDS.

To investigate the effects of the proportions of aberrant cells on the neural manifolds, we simulated the same models with varying percentages of normal, hyperactive ($\alpha$) and hypoactive ($\beta$) cells.  As expected, the hyperactivity of cells close to amyloid plaques (which tends to add noise) distorts the neural manifolds more than the hypoactive cells (which is more akin to having a decreased number of neurons in the manifold), resulting in less clustering of neural states in the manifold space. This is consistent with their effects on performance in terms of the given manifold measures (Figure \ref{fig:disease}f-g, \ref{fig:disease}i-j, \ref{fig:disease}l-m). 

We suggest that on this basis, NML techniques exhibit potential for helping to improve our understanding of network-level disruption in disease phenotypes. 

\section{Discussion}
\label{S:discussion}

\subsection{Comparison of neural manifold algorithms}
\label{S:manifolddiscussion}

\noindent With the increase in the numbers of simultaneously recorded neurons facilitated by emerging recording technologies, the demand for methods that enable reliable and comprehensively informative population-level analysis also increases. Multivariate analyses of large neuronal population activity thus require the use of more efficient computational models and algorithms that are geared towards finding the fewest possible population-level features that best describe the coordinated activity of such large neuronal populations. These features must be interpretable and open to parameterization that enables inherent neural mechanisms to be described (as discussed in the next section). 

In this review, we give insights into how NML facilitates accurate quantitative characterizations of the dynamics of large neuronal populations in both healthy and disease states. However, assessing exactly how reliable and informative NML methods are remains a challenge and is open to interpretation. An ideal NML method must be able to adapt to different datasets that span functionally different brain regions, behavioral tasks, and number of neurons being analyzed,  while considering different noise levels, timescales, etc. 

In many studies, linear NML methods (such as PCA and classical MDS) have been preferred, and are widely used throughout the literature because of computational simplicity and clear interpretation. However, these methods have not been sufficient when the geometry of the neural representation is highly non-linear, which might for instance be induced by high behavioral task complexity \citep{jazayeri2021interpreting} (see Figure \ref{fig:prefrontal}). As shown in \citep{rubin2019revealing}, low SNR can also affect the quality of some linear NML embeddings. They compared PCA and LEM algorithms and revealed that LEM required fewer neurons and lower firing rates for accurate estimation of the internal structure. To address the challenges of noisy data, many solutions have been suggested, including the development of novel linear algorithms, such as dPCA \citep{kobak2016dpca}, cvPCA \citep{stringer2019high}, and mTDR \citep{Aoi2018mTDR}. These methods enable greater discrimination between specific clusters of neural activity, but they do rely heavily on neural responses acquired during multiple trials of time-locked behavior. However, these may not always be available, and whilst it is possible to tweak certain algorithmic parameters or fit them to manipulated forms of the data (e.g. through time-warping), the end result is an increase in computational complexity and possibly harder to interpret results. 

Conversely, non-linear methods, in general, trend towards better generalization across datasets. This is particularly true for UMAP and t-SNE (see Figures \ref{fig:CA1} and \ref{fig:motor}, \ref{fig:prefrontal}). To discriminate clusters of neural activity, UMAP and t-SNE have been shown to be the most powerful non-linear NML techniques. However, they might not always yield an embedding that is representative of the true high-dimensional geometry. These issues are particularly relevant for inherently very high-dimensional data, such as genomics data \citep{Chari2021TheGenomics}. In fact, it is debated whether a low number of latent dimensions are enough to drive neural population activity and explain its variability across a range of circumstances such as brain region and task \citep{Chaudhuri2019TheSleep,rubin2019revealing,nieh2021geometry, Churchland2012NeuralReaching, Gallego2020Long-termBehavior}, or if higher dimensionality are required \citep{stringer2019high,humphries2020strong}.

Finally, a key consideration when selecting any NML method is whether its function is for neural data visualization or for structural discovery and/or quantification. For instance, many of the non-linear algorithms described above, such as UMAP, t-SNE have excellent capabilities in visualizing high-dimensional data and provide an interpretable global structure. Similarly, VAEs present themselves as great candidates for discriminating global clusters. However, obtaining a favorable global structure can often be at the expense of a loss to local structure \citep{Chari2021TheGenomics} and therefore can distort any downstream structural quantification.  To address these shortcomings one can use algorithms that either provide a linear solution such as PCA-based methods \citep{gardner2022toroidal} or that focus on local structure such as LEM \citep{rubin2019revealing}. ANNs may add further possibilities, with methods such as CEBRA tackling model identifiability and generalization \citep{schneider2022learnable}.

One message that can be taken away from our comparison of neural manifold approaches is that linear methods do provide substantial insight over a wide range of problem domains (often consistent with what is provided by non-linear methods), and even in areas where they fail to correctly capture the low-dimensional topological structure of the manifold, they can still provide insight verifiable through decoding of behavior. 

\subsection{Parameterization of neural manifolds}
\label{S:parameterization}

\noindent  A key issue for NML approaches is that after the neural manifold has been determined, it normally needs to be analyzed further to answer a given scientific question. This process, known as parameterization, would ideally involve extraction of the equations of the manifold itself, however, out of feasibility may also be based on a more limited quantification of manifold properties. Depending on the type of neural data at hand and the scientific hypothesis to be tested, it's important to select the most suited and informative NML algorithm and parameterization measures. One challenge that has to be kept in mind is that the resulting manifolds, although often visualized in two or three dimensions for convenience, frequently involve many more dimensions (e.g. even a low-dimensional neural manifold from a recording from many hundreds of neurons might involve tens of dimensions), and many approaches that may work well in two or three dimensions do not necessarily scale straightforwardly to higher-dimensional representations.

Where transitions in the dynamics of neural states are observed over time, one approach that is useful is  recurrence analysis \citep{marwan2007recurrence, wallot2019multidimensional}. Recurrence analysis is a non-linear data analysis that is based on the study of phase space trajectories, where a point or element in phase space represents possibles states of the system. It is a powerful tool in characterizing the behavior of a dynamical system, especially when observing how states change over time. A recurrence plot measures and visualizes the recurrences of a trajectory of a system in phase space. Note that although the recurrent plot is inherently two-dimensional, recurrence analysis can describe dynamics in an arbitrary dimensional space - a key feature of this technique. As the respective phase spaces of two systems change, recurrence plots allow quantification of their interaction and tell us when similar states of the underlying system occur. The time evolution of trajectories can be observed in the patterns depicted in the recurrence plots, and these patterns are associated with a specific behavior of the system, such as cyclicity and similarity of the evolution of states at different epochs. Different measures can be used to quantifying such recurrences based on the structure of the recurrence plot, including recurrence rate, determinism (predictibility), divergence, laminarity, and Shannon entropy \citep{webber2015recurrence,wallot2019multidimensional}. These measures might be especially useful when observing neuronal responses to stimuli over repeated trials under different conditions (such as healthy and diseased states). 

Another way to compare neural manifolds under different conditions is to directly quantify the similarity of trajectories in the neural manifold space \citep{cleasby2019using, ding2008querying,alt2009computational}. The most commonly used measures of trajectory similarity are Fr{\'e}chet distance (FD) \citep{frechet1906quelques,besse2015review}, nearest neighbor distance (NND) \citep{clark1954distance,freeman2011group}, longest common subsequence (LCSS) \citep{vlachos2002discovering}, edit distance for real sequences (EDR) \citep{chen2004marriage,chen2005robust} and dynamic time warping (DTW) \citep{toohey2015similaritymeasures,long2013review}. Computing such measures may allow us to characterize the differences (or similarities) among neural manifolds obtained from different models. This type of analysis may be especially useful when comparing models in health and disease states across age groups, or when comparing neural manifold representations of the same model across different environments and conditions. 

Another useful measure is manifold trajectory tangling \citep{russo2018motor}. Tangling is a simple way of determining whether a particular trajectory could have been generated by a smooth dynamical flow field; under smooth dynamics, neural trajectories should not be tangled. Two neural states are thus tangled if they are nearby but associated with different trajectory directions. High tangling implies either that the system must rely on external commands rather than internal dynamics, or that the system is flirting with instability \citep{russo2018motor}. Thus, tangling can be compared across all times and/or conditions. It is especially useful when characterizing neural dynamics over the course of learning or development. For example, it may be interesting to examine whether neural circuits adopt network trajectories that are increasingly less tangled when learning a new skill and with increasing performance. Conversely, it may also be interesting to investigate whether pathological conditions may be associated with increased tangling during learning. 

In cases where we characterize task-specific responses, e.g., during repetitive, trial-structured motor movements, determining whether these neural responses are consistent with the hypothesised computation is important. Hence, the consistency in the geometry of the population response and/or the time-evolving trajectory of activity in neural state space (i.e., manifold space) must be examined. For such task-specific trials, manifold trajectories are considered consistent if the trajectory segments trace the same path and do not diverge. One measure that can be used for such analysis is manifold trajectory divergence \citep{russo2020neural}. In contrast to manifold trajectory tangling, which assesses whether trajectories are consistent with a locally smooth flow field, manifold trajectory divergence assesses whether similar paths eventually separate, smoothly or otherwise. A trajectory can have low tangling but high divergence, or vice versa. Multiple features can contribute to divergence, including ramping activity, cycle-specific responses, and different subspaces on different cycles. Thus, manifold trajectory divergence provides a useful summary of a computationally relevant property, regardless of the specifics of how it was achieved \citep{russo2020neural}.

Lastly, topological data analysis (TDA), which comprises methods such as persistent cohomology, aim to rigorously characterize the topological structure of the data. These methods are quite sensitive to noise levels and might not be adequate to investigate functionally complex neural systems, but they succeed in revealing the topology of simpler systems such as the head direction, grid cell systems and linear track representation \citep{Chaudhuri2019TheSleep, gardner2022toroidal, rubin2019revealing}. Similarly, Spline Parameterization for Unsupervised Decoding (SPUD) is a powerful method to characterise the manifold dynamics, which stems from a rigorous analysis of the manifold topological structure \citep{Chaudhuri2019TheSleep}.

An increasing body of work using large-scale neural recording technologies is pointing towards the viewpoint that neural manifolds during spontaneous activity may be relatively high dimensional \citep[reviewed in][]{avitan2022not}, in comparison to the low-dimensional picture of spontaneous activity that had emerged from earlier recording technologies such as single-unit extracellular electrophysiology and functional magnetic resonance imaging. This of course provokes the question: at what dimensionality do neural manifold approaches cease to become useful? Notably, the datasets we have used as illustrations in this survey paper are (reflecting the field) intrinsically relatively low-dimensional in nature, ranging from a mouse running around a constrained environment to a monkey moving its arm in a fixed set of trajectories. Free-ranging, real-world behaviour is obviously higher dimensional in nature. Do manifold approaches cease to become useful in such cases? We would argue not, as long as the dimensionality of the manifold is much lower than the dimensionality of the neural recording (essentially determined by the number of cells that can be recorded, and which can exceed 10,000 with recently introduced technologies). Certainly two and three-dimensional visualisations may fail to capture interesting aspects of the data (which may in many cases be true already for 10-dimensional datasets). However, as we have pointed out in this section, there are many approaches for analysis of the resulting manifolds, such as recurrence plots, trajectory similarity measures, and manifold trajectory tangling, that work for higher dimensional manifolds. How these properties relate back to the predictions made by computational models of brain function is so far less clear. This area is ripe for further theoretical work.

\subsection{Is neural manifold learning useful for understanding neurological disorders?}
\label{S:manifolddisorders}

\noindent Neural manifold learning offers the prospect of aiding in our understanding of circuit neuropathologies. For instance, in mouse models of Alzheimer's disease, amyloid or tau pathologies result in changes in cortical circuitry, which are particularly evident in the hippocampus and connected structures. However, the effect of these pathologies on the dynamics of neural circuits involved in spatial and working memory at the network level are still not well understood. By analyzing the changes in the geometry of population responses and the time-evolving trajectory of activity of the associated hippocampal-cortical circuits, we can compare the neural manifolds recovered from groups of mice of different ages and/or with different health states. In particular, we can use recurrence analysis to determine whether neural states recur upon presentation of the same stimuli over repeated trials at different times across different ages of the models. Neural manifold similarity, tangling and divergence measures may also be computed to evaluate the differences (or similarity) of movement trajectories in the neural manifold space. Together, these measures can be used to compare the manifold dynamics across different age groups in both healthy and disease states. 

In this review, we have demonstrated that neural manifold learning methods provide a powerful toolbox for understanding how populations of neurons coordinate in representing behaviorally relevant information. While the cellular and molecular pathologies underlying a variety of neurological disorders such as Alzheimer's Disease, Parkinson's Disease and Frontotemporal Dementia are at least beginning to be well understood, how they translate into network dysfunction and thus, into cognitive and behavioral deficits is not. Neural manifold learning techniques, in combination with new experimental technologies allowing us to record the activity of many thousands of neurons simultaneously during behavioral tasks, potentially may allow us to assay the dynamics and Gestalt representations underlying cognition and behavior at the level of entire neural circuits. This could in turn lead to improved understanding of disease processes and more sensitive tests of therapeutic effect.

\bmhead{Acknowledgments}
We thank Krishna V Shenoy and Byron Yu for their permission to use the macaque motor cortex dataset and Alon Rubin for permission to use the mouse ACC dataset. We thank Elena Faillace for useful comments on a previous version of this manuscript.

\bmhead{Funding}
This work was funded by studentships from the EPSRC CDT in Neurotechnology for Life and Health (EPSRC EP/L016737/1) to S.P. and G.P.G., by Wellcome Trust Award 22152/Z/20/Z (S.R.S. and M.A.G.), by Wellcome Trust Award 207481/Z/17/Z (salary support to RM-H from R.G.M. Morris) and by Mrs Anne Uren and The Michael Uren Foundation. 

\bmhead{Rights retention}
This research was funded in whole, or in part, by the Wellcome Trust [22152/Z/20/Z,207481/Z/17/Z]. For the purpose of Open Access, the author has applied a CC BY public copyright licence to any Author Accepted Manuscript version arising from this submission.

\bmhead{Conflict of interest/Competing interests} The authors declare that there are no conflicts of interest.

\bmhead{Code availability} 
Code to generate some of the figures in this paper is available at: \url{https://github.com/schultzlab/Neural_Manifolds}

\bmhead{Authors' contributions}
Rufus Mitchell-Heggs co-led and managed the project wrote and reviewed the main manuscript text, designed code for figures 3-5 and co-prepared figures 1,3,4,5,6.
Seigfred Prado wrote and reviewed the main manuscript text, designed code for figures 6 and co-prepared figures 1 and 5.
Giuseppe P. Gava wrote and reviewed the main manuscript text and designed the majority of manifold learning code for figures 3-6 and created figure 2.
Mary Ann Go contributed experimental data used for figure 3 and reviewed the main manuscript.
Simon R. Schultz co-led and managed the project, wrote and reviewed the main manuscript.

\bibliography{sn-bibliography}{99}
\setlength{\bibsep}{0pt plus 0.3ex}

\end{document}